\documentclass[twocolumn,prl,showpacs,amsmath,amssymb]{revtex4}
\usepackage{graphicx}
\usepackage{amsfonts}

\def\be{\begin{equation}}
\def\ee{\end{equation}}
\def\bea{\begin{eqnarray}}
\def\eea{\end{eqnarray}}
\def\bma{\begin{mathletters}}
\def\ema{\end{mathletters}}

\def\0{\overline{0}}

\def\q0{\underline{0}}

\def\id{{\mathbb I}}

\def\tr{\mbox{tr}}
\def\one{\leavevmode\hbox{\small1\normalsize\kern-.33em1}}

\def\bra#1{\langle#1|} \def\ket#1{|#1\rangle}

\def\proj#1{\ket{#1}\!\bra{#1}}

\begin{document}

\title{Pure state estimation and the characterization of entanglement}
\author{Miguel Navascu\'es}

\affiliation{ICFO-Institut de Ciencies Fotoniques, Mediterranean
Technology Park, 08860 Castelldefels (Barcelona), Spain}

\begin{abstract}
A connection between the state estimation problem and the
separability problem is noticed and exploited to find efficient
numerical algorithms to solve the first one. Based on these ideas,
we also derive a systematic method to obtain upper bounds on the
maximum local fidelity when the states are distributed among
several distant parties.
\end{abstract}

\maketitle

The estimation of specific properties of a quantum system subject
to observation was one of the very first problems posed in Quantum
Information Theory \cite{state}. This problem has no analog in the
classical case, where all measurements commute and a sufficiently
subtle observer (a Maxwell demon) is able to make complete
tomography of a system without introducing any perturbations in
it. In the quantum case, measurements are irreversible and
invariably destroy information, and thus we have to design and
optimize our measurement strategy according to some figure of
merit, that is, we have to decide ``what to see" before we can
extract conclusions from ``what we see". This figure of merit can
be, for instance, the mean variance between our estimator and the
quantity we want to determine \cite{yuen}, or the mutual
information between the outcome of our measurement and the
parameters that describe the possible set of states
\cite{holevobound}. When the figure of merit is related to the
overlap between two quantum states, then we are dealing with a
state estimation problem \cite{popescu}. In its most usual form,
this consists on estimating the whole wave function of an unknown
state when only a limited number of copies of this state is
available. Initially considered as a purely theoretical problem
interesting by itself, it has immediate applications in the
calibration of quantum optical devices, where complete tomography
is impractical, or even impossible (like when we deal with
continuous variable systems, for example). Moreover, a variant of
state estimation, namely, state discrimination, has recently being
used in Quantum Computation to find efficient quantum algorithms
to solve the Hidden Subgroup Problem \cite{Ip}.

In a \emph{general} state estimation scenario, a source produces
with probability $p_i$ a quantum state $\Psi_i$ that is encoded
afterwards into another quantum state $\Psi'_i$, to which we have
full access. Our duty is to measure our given state and thus
obtain a classical value $x$ that we may use to make a guess on
the original state $\Psi_i$, which we will assume to be pure along
this article. For abbreviation, we will label this type of
problems by $(p_i,\Psi_i,\Psi_i')$. Games belonging to this family
appear quite often in Quantum Information Theory. For example, we
could think of quantum tomography as a protocol in which a source
produces a state $\Psi_i$ and encodes it into
$\Psi'_i=\Psi_i^{\otimes N}$. Analogously, state discrimination
can be easily proven to be equivalent to a state estimation
problem where the source produces orthogonal states $\{\proj{i}\}$
and encodes them into non-orthogonal states $\{\Psi_i\}$. A usual
figure of merit to quantify the knowledge on $\Psi_i$ provided by
$x$ is the so called \emph{fidelity}. The idea is to prepare a
state $\phi_x$ as a guess and then compare this state to the one
that was originally encoded by the source just by computing the
overlap between the two: $\tr(\phi_x\Psi_i)$ (remember that we
assume the states $\{\Psi_i\}_i$ to be pure). If we denote by
$M_x$ each of the Positive Operator Value Measure (POVM) elements
that mathematically describe our measurement of the state
$\Psi_i'$, then the efficiency of our \emph{strategy}
$(M_x,\phi_x)$ will then be determined by the \emph{average
fidelity} $f$:

\begin{equation}
0\leq f\equiv\sum_{i,x}p_i\tr(\Psi_i'M_x)\tr(\phi_x\Psi_i)\leq 1.
\end{equation}

The state estimation problem consists on determining $F$, defined
as the maximum fidelity among all possible strategies
$(M_x,\phi_x)$.

Up to now, we have been considering \emph{global state
estimation}. But there are many interesting situations in Quantum
Information Theory where the encoded states $\{\Psi'_i\}_i$ are
distributed between two or more distant parties. In this
situation, the parties will have to agree on a strategy based on
Local Operations and Classical Communication (LOCC) that allows
them to extract enough information about the sent states so that
they can make a good guess on the state produced by the source.
Then, we will be interested in determining the \emph{local
fidelity} $F_L$:

\begin{eqnarray}
&F_L=\sup\sum_{i,x}p_i\tr(\Psi_i'(M_x)_{AB})\tr(\phi_x\Psi_i),
\end{eqnarray}

\noindent where the $(M_x)_{AB}$ correspond to the POVM elements
of a measurement accessible via LOCC.

\noindent The main result of this paper is the following:

\noindent\textbf{Proposition 1}

For any global state estimation problem with solution $F$, there
exists a sequence of real numbers $F^{(1)},F^{(2)},F^{(3)},...$,
where each of the elements can be computed efficiently, and such
that $F^{(1)}\geq F^{(2)}\geq F^{(3)}\geq...\geq F$ and
$\lim_{n\to \infty}F^{(n)}=F$.

This proposition can be adapted to deal with the problem of local
state estimation, and the result is

\noindent\textbf{Proposition 2}

For any local state estimation problem with solution $F_L$, there
exists a sequence of real numbers
$F_S^{(1)},F_S^{(2)},F_S^{(3)},...$, where each of the elements
can be computed efficiently, and such that $F_S^{(1)}\geq
F_S^{(2)}\geq F_S^{(3)}\geq...$ and $\lim_{n\to
\infty}F_S^{(n)}=F_S\geq F_L$, where $F_S$ corresponds to the
\emph{separable fidelity}, i.e., the maximum attainable fidelity
when the POVM is separable with respect to the $k$ parties.

The structure of this paper is as follows: first, we will show the
connection between the state estimation problem and the
separability problem and use it to proof Proposition 1. An
isotropic distribution of states of arbitrary dimension will be
considered to illustrate the efficiency of the resulting numerical
tools. Proposition 2 will follow naturally from Proposition 1, and
the problem of local state estimation of an arbitrary probability
distribution of Bell states will be solved as an example. Finally,
we will expose our conclusions.

For any state estimation problem $(p_i,\Psi_i,\Psi_i')$ and any
state estimation strategy $(M_x,\phi_x)$, the corresponding
fidelity $f$ can always be expressed as

\begin{equation}
f=\tr(\rho_{AB}\Lambda_{AB}),
\end{equation}

\noindent with $\rho_{AB}=\sum_i p_i \Psi'_i\otimes\Psi_i$,
$\Lambda_{AB}=\sum_x M_x\otimes\phi_x$. Note that, while
$\rho_{AB}$ is a separable quantum state fixed by the state
estimation problem, $\Lambda_{AB}$ only depends on our
measure-and-prepare strategy. It is immediate to see that for
$\Lambda_{AB}$ to describe the action of a strategy over
$\rho_{AB}$, $\Lambda_{AB}$ must be a separable positive
semidefinite operator. Moreover, $\tr_B\Lambda_{AB}=\sum_x
M_x={\mathbb I}_A$. Conversely, any separable positive
semidefinite operator $\Lambda_{AB}$ with partial trace equal to
the identity can be made to correspond to a state estimation
strategy. Thus, the state estimation problem can be reformulated
in this way:

\begin{equation}
F=\sup\{\tr(\Lambda_{AB}\rho_{AB}):\Lambda_{AB} \mbox{
sep},\tr_B\Lambda_{AB}={\mathbb I}\}. \label{reformu}
\end{equation}

To see how this approach works, consider the problem of estimating
pure states that are distributed according to an isotropic
probability density, i.e., we are dealing with
$(dU,U\proj{0}U^\dagger,U\proj{0}U^\dagger)$, where $U$ is a
unitary operator and $dU$ denotes the Haar measure of the unitary
group $SU(d)$. Then,

\begin{equation}
\rho_{AB}=\int U\proj{0}U^\dagger\otimes U\proj{0}U^\dagger
dU=\frac{1}{d(d+1)}(\id + V), \label{roer}
\end{equation}

\noindent where $V$ denotes the flip operator. Now, take
$\Lambda_{AB}$ be an operator associated to a possible state
estimation strategy for this problem. It is easy to see that
$f=\tr(\rho_{AB}\Lambda_{AB})=\tr(\rho_{AB}\tilde{\Lambda}_{AB})$,
where $\tilde{\Lambda}_{AB}=\int U\otimes
U\Lambda_{AB}U^\dagger\otimes U^\dagger dU$. Note that $\tilde{\Lambda}_{AB}/d$ is a
separable state and
$\tr_B(\tilde{\Lambda}_{AB})=\id$, i.e., $\tilde{\Lambda}_{AB}$ corresponds to a strategy.

The above argument shows that we can take $\Lambda_{AB}/d$ to be
a Werner state \cite{werner}. Werner states are a monoparametrical
family of states whose partial trace is proportional to the
identity and whose separability regions are identified. Following
\cite{werner}, we can write $\Lambda_{AB}$ as
$\Lambda_{AB}=\frac{1}{d(d^2-1)}\left[(d-t)\id+(dt-1)V\right]$,
where $-1\leq t\leq 1$ guarantees that $\Lambda_{AB}$ is positive
semidefinite and $t\geq 0$ implies that it is separable. A direct
calculation shows that

\begin{equation}
\tr(\rho_{AB}\Lambda_{AB})=\frac{1+t}{d+1}. \label{wernito}
\end{equation}

Thus, we get that $F=\frac{2}{d+1}$, recovering the results of
Bru$\beta$ and Macchiavello \cite{bruss}.

To solve the above problem we played with the advantage that the
separability properties of Werner states are well known. However,
most state estimation problems do not have the symmetries of
(\ref{roer}). How could we exploit the reformulation of the
problem given by (\ref{reformu}) in order to address the general
case? The approach we chose makes use of a characterization of the
set of all separable states due to Doherty et al. \cite{doherty}.
This characterization is based on the notion of \emph{PPT
symmetric extensions}.

A \emph{PPT symmetric extension $\bar{\Lambda}^{(n)}_{AB}$ to $n$
copies of party $A$} of a bipartite quantum state
$\bar{\Lambda}_{AB}$ is a positive semidefinite operator
$\bar{\Lambda}^{(n)}_{AB}\in{\cal B}({\cal H}_A^{\otimes n}\otimes
{\cal H}_B)$ such that

\noindent 1) $\bar{\Lambda}^{(n)}_{AB}$ is symmetric with respect
to the interchange of any two copies of system $A$.

\noindent 2) $\bar{\Lambda}^{(n)}_{AB}$ returns the state
$\bar{\Lambda}_{AB}$ when $n-1$ copies of system $A$ are traced
out.

\noindent 3) $\bar{\Lambda}^{(n)}_{AB}$ remains positive under all
possible partial transpositions.

We will define $S^{(n)}$ as the set of all bipartite operators
$\bar{\Lambda}_{AB}$ that admit a PPT symmetric extension to $n$
copies of system $A$.

If we call $S$ to the set of all separable states, it is not
difficult to prove that any element of $S$ must belong to
$S^{(n)}$ for any $n$ \cite{demo}. This observation immediately
leads us to an infinite set of necessary conditions for a state to
be separable. Moreover, this set of conditions constitutes a
hierarchy, since it follows from the definition of PPT symmetric
extension that any state belonging to $S^{(n+1)}$ has necessarily
to belong to $S^{(n)}$. What it is not so obvious is that this
hierarchy is complete. That is, any quantum state belonging to
$S^{(n)}$ for all $n$ can be proven to be separable
\cite{doherty}.

Let $d_A$ be the dimension of ${\cal H}_A$. Consider now the
following problem:

\begin{equation}
F^{(n)}=\sup\{\tr(\Lambda_{AB}\rho_{AB}):\Lambda_{AB}/d_A \in
S^{(n)},\tr_B\Lambda_{AB}={\mathbb I}\}. \label{ideacentral}
\end{equation}

From what we have seen, it is evident that $F^{(1)}\geq
F^{(2)}\geq F^{(3)}\geq...\geq F$ and, moreover,
$\lim_{n\to\infty} F^{(n)}=F$. The advantage of this approach is
that each of the bounds $F^{(n)}$ can be computed efficiently
using semidefinite programming.

Semidefinite programming is a branch of numerical analysis that is
concerned with the following optimization problem:
\begin{eqnarray}\label{primal}
&\mbox{minimize } \vec{c^T}\cdot \vec{x}\nonumber\\
&\mbox{subject to } F(\vec{x})\equiv F_0+\sum_{i=1}^n x_iF_i\geq
0,  \end{eqnarray} where $\{F_i\}_{i=0}^n$ are $N\times N$
matrices. This is known as the \emph{primal problem} and its
solution is usually denoted by $p^*$. For each primal problem
there is an associated \emph{dual problem}, of the form

\begin{eqnarray}
&d^*\equiv\mbox{maximize } -\mbox{tr}(F_0Z)\nonumber\\
&\mbox{subject to } Z\geq
0,\mbox{tr}F_iZ=c_i,i=1,...,n,\label{dual}
\end{eqnarray}

\noindent that can also be treated a semidefinite program. It can
be seen that $d^*\leq p^*$ \cite{sdp}. Moreover, a sufficient
condition to assure that $d^*=p^*$ is the existence of a
\emph{strict primal feasible} point, that is, a vector $\vec{x}$
such that $F(\vec{x})>0$. Note that this condition holds in our
case, for it is straightforward that the matrix
$\bar{\Lambda}^{(n)}_{AB}\equiv (\id_A/d_A)^{\otimes
n}\otimes\id_B/d_B>0$ corresponds to a PPT symmetric extension to
$n$ copies of system $A$ of the quantum state
$\Lambda_{AB}/d_A\equiv (\id_A/d_A)\otimes\id_B/d_B$, with
$\tr_B(\Lambda_{AB})=\id$.

The scheme to solve (\ref{ideacentral}) would then be to run
algorithms that try to solve both the dual and the primal problem
at the same time. If, after some iterations, our computer returns
the points $\vec{x},Z$, we know that the solution of our problem
lies somewhere in the middle, i.e., $\mbox{tr}(ZF_0)\leq p^*\leq
\vec{c}^T\cdot \vec{x}$. Because semidefinite programs belong to
the $P$ complexity class and the matrices involved in the
calculation of each bound $F^{(n)}$ increase in size as
$d_A^nd_B$, for fixed $n$, $F^{(n)}$ can be calculated
efficiently, as promised. We have proven Proposition 1.

Although Proposition 1 only guarantees the asymptotic convergence
of the series $(F^{(n)})$ to the optimal fidelity, for some
specific problems we may find the value of $F$ after a few
iterations. Returning to the task of estimating isotropically
distributed quantum states, because the maximum of (\ref{wernito})
when $\Lambda_{AB}$ is only demanded to be positive semidefinite
is the same as the maximum among the set of all separable
operators, it follows that $F^{(1)}=F$, i.e., if we had performed
a numerical optimization, the first upper bound would have
coincided with the optimal fidelity. Also, in case the original
and encoded states of the protocol are elements of a Hilbert space
of dimension 2, the PPT criterion is sufficient to guarantee the
separability of $\Lambda_{AB}$ \cite{horodecki}, and thus, for all
these problems, $F=F^{(1)}$.

These ideas can be extended to obtain upper bounds on the solution
of more complicated problems. Suppose that the given state
$\Psi'_i$ is a bipartite state that is distributed by a source to
two spatially separated parties, Alice and Bob. Then, Alice and
Bob would have to agree on a protocol based on local operations
and classical communication that allowed them to obtain classical
information about the original state $\Psi_i$.

This problem can be reformulated in a similar fashion as
(\ref{reformu})

\begin{equation}
F_L=\sup\{\tr(\Lambda_{ABC}\rho_{ABC}):\Lambda_{ABC} \mbox{
LOCC}\}, \label{reformu2}
\end{equation}

\noindent where $\rho_{ABC}=\sum_i
p_i(\Psi'_i)_{AB}\otimes\Psi_i$, and $\Lambda_{ABC}$ denotes an
element of the set of all strategies that are accessible to Alice
and Bob using LOCC. Unfortunately, this problem cannot be solved
directly, for up to now no one has been able to characterize the
set of all POVMs accessible by LOCC. However, we know that such a
set is contained in the set of all separable POVMs, and thus we
may conform with an upper bound on $F_L$, namely, $F_S$, the
\emph{separable fidelity}, which we could define as

\begin{equation}
F_S=\sup\{\tr(\Lambda_{ABC}\rho_{ABC}):\Lambda_{ABC} \mbox{
trisep},\tr_C\Lambda_{ABC}={\mathbb I}\}. \label{reformu3}
\end{equation}

It is known that $F_L$ and $F_S$ do not coincide in general. For
example, the separable fidelity of a uniform distribution of the
domino states \cite{bennett} would be 1, whereas $F_L$ can be
proven to be strictly smaller than 1. However, sometimes $F_S$ can
be a reasonable approximation to $F_L$, as we will soon see.

There exists a complete characterization of the set of all
$k$-separable states, also due to Doherty et al. \cite{doherty2},
and also implementable using semidefinite programming packages.
This characterization is just a generalization of the previous
one. In this case, we would demand from any triseparable
(normalized) state $\bar{\Lambda}_{ABC}$ that it admits a PPT
symmetric extension to $n$ copies of systems $A$ and $B$, for any
$n$. Analogously to the global state estimation case, we would
obtain a sequence of upper bounds $F_S^{(1)}\geq
F_S^{(2)}\geq...\geq F_L$. However, this time
$\lim_{n\to\infty}F_S^{(n)} = F_S\geq F_L$. This scheme can be
easily generalized to any number of distant parties. Proposition 2
has been proven.

To see the usefulness of this method, consider the particular
example where Alice and Bob are distributed one of the four Bell
states with different probabilities, and that each of these states
encodes the very same Bell state, that is, if we call
$\{\psi_i\}_{i=1}^4$ to the four Bell states, we would be dealing
with the problem $(p_i,\psi_i,(\psi_i)_{AB})$. Take

\begin{eqnarray}
&\psi_{1,2}=\frac{1}{2}(\ket{00}\pm\ket{11})(\bra{00}\pm\bra{11}),\nonumber\\
&\psi_{3,4}=\frac{1}{2}(\ket{01}\pm\ket{10})(\bra{01}\pm\bra{10}).
\end{eqnarray}

We are going to calculate $F_S^{(1)}$. The corresponding primal
problem is

\begin{eqnarray}
&\mbox{maximize } \tr(\rho_{ABC}\Lambda_{ABC})\nonumber\\
&\mbox{subject to } \tr_C(\Lambda_{ABC})={\mathbb I},\nonumber\\
&\Lambda_{ABC},\Lambda^{T_A}_{ABC},\Lambda^{T_B}_{ABC},\Lambda^{T_C}_{ABC}\geq
0,
\end{eqnarray}

\noindent with $\rho_{ABC}=\sum_i
p_i(\psi_i)_{AB}\otimes(\psi_i)_C$.

The dual of this problem (module some simplifications) is

\begin{eqnarray}
&\mbox{minimize } \tr(\tilde{\rho})\nonumber\\
&\mbox{subject to }A,B,C\geq 0,\nonumber\\
&\tilde{\rho}\otimes{\mathbb
I}_C-A^{T_A}-B^{T_B}-C^{T_C}-\rho_{ABC}\geq 0.
\end{eqnarray}

From our previous discussion about semidefinite programing, it
follows that any feasible point of the dual of a semidefinite
program provides an (in this case) upper bound on the solution of
the primal problem. The fact that $\psi_i^{T_A}={\mathbb
I}-2\psi_{5-i}$ suggests that we may try to solve the dual problem
using the following ansatz:

\begin{eqnarray}
&A=\sum_i \frac{\lambda_i}{2}\psi_{5-i}\otimes\psi_i,B=C=0,\nonumber\\
&\tilde{\rho}=\sum_j\mu_j\psi_j.
\end{eqnarray}

$A\geq 0$ implies that $\lambda_i\geq 0,\forall i$. Analogously,
$\tilde{\rho}\otimes{\mathbb I}_C-A^T_A-\rho_{ABC}\geq 0$ implies
that

\begin{eqnarray}
&\mu_i\geq p_i-\frac{\lambda_i}{2}\nonumber\\
&\mu_i\geq\frac{\lambda_j}{2},\forall j\not=i.
\end{eqnarray}

What we have to minimize is the quantity $\bar{f}=\sum_j\mu_j$. If
we fix the $\lambda$s, the minimum value of each $\mu_i$ will be
$\mu_i=\max(\{\frac{\lambda_j}{2}:j\not=i\},p_i-\frac{\lambda_i}{2})$.
Therefore, $\bar{f}=\sum_i
\max(\{\frac{\lambda_j}{2}:j\not=i\},p_i-\frac{\lambda_i}{2})$.
Because of the symmetry of the problem, we can suppose that
$\lambda_1\geq\lambda_2$ are the two greatest $\lambda$s. Then,
$\bar{f}=\max(\frac{\lambda_2}{2},p_1-\frac{\lambda_1}{2})+\sum_{i=2}^4
\max(\frac{\lambda_1}{2},p_i-\frac{\lambda_i}{2})$. This quantity
can only be made smaller if we take $\lambda_{3,4}=\lambda_2$, and
so we can assume
$\bar{f}=\max(\frac{\lambda_2}{2},p_1-\frac{\lambda_1}{2})+\sum_{i=2}^4
\max(\frac{\lambda_1}{2},p_i-\frac{\lambda_2}{2})$.

On the other hand, for any choice of $\lambda_1,\lambda_2$, it can
be proven that $\bar{f}$ becomes strictly smaller if we take
$\lambda'_1=\lambda'_2=(\lambda_1+\lambda_2)/2\equiv\lambda$.
Finally, $\bar{f}=\sum_{i=1}^4
\max(\frac{\lambda}{2},p_i-\frac{\lambda}{2})$. If we order the
probabilities such that $p_a\geq p_b\geq p_c\geq p_d$, it is
immediate to check that the guess $\lambda=p_c$ provides the
minimum value $\bar{f}=p_a+p_b$. Therefore, $F_L\leq F^{(1)}_S\leq
p_a+p_b$.

But such fidelity is actually attainable! All Alice and Bob have
to do is to measure either in the computational or in the $X$
basis each and prepare the Bell states with greatest probability
that are compatible with their measurements. For example, in the
case where $p_a=p_1,p_b=p_2$, Alice and Bob should measure in the
$X$ basis. If their results are correlated, they should prepare
the state $\psi_1$. Otherwise, they should prepare the state
$\psi_2$. It is straightforward to see that this strategy attains
a fidelity $f_L=p_1+p_2$.

Incidentally, because the Bell states are orthogonal to each
other, this state estimation problem is equivalent to the state
discrimination problem $(p_i,\proj{i},\Psi_i)$, that is, $F_L$
also corresponds to the maximum probability of determining which
Bell state was sent to Alice and Bob.

In conclusion, we have shown a systematic method to numerically
solve the state estimation problem, and we have seen how this
approach can be slightly modified to attack the local state
estimation problem. Semidefinite programming had already been used
in the resolution of state discrimination problems
\cite{fiurasek,Ip}, but, to the author's knowledge, this is the
first time such a mathematical tool is proposed to deal with the
general state estimation scenario.

Along this paper, we have been tacitly assuming that we were
dealing with state estimation problems in finite dimensions. We
have ideas about how to treat the infinite dimensional case, that
will be developed in future publications.

\emph{Acknowledgements:} We thank Antonio Ac\'{\i}n for useful
discussions and the Fundaci\'on Ram\'on Areces for financial
support.


\begin{thebibliography}{99}
\bibitem{demo}
If $\bar{\Lambda}_{AB}$ is a separable state, then there exists a
decomposition $\bar{\Lambda}_{AB}=\sum_i
p_i\Lambda_i\otimes\Lambda'_i$. It is immediate to check that the
state $\sum_i p_i\Lambda_i^{\otimes n}\otimes\Lambda'_i$ is a PPT
symmetric extension of $\bar{\Lambda}_{AB}$ to $n$ copies of party
$A$.
\end{thebibliography}
\end{document}